\documentclass[bibyear]{aa}

\usepackage{graphicx}
\usepackage{txfonts}
\usepackage{dblfloatfix}
\usepackage{natbib}
\usepackage{hyperref}
\usepackage{knitting}
\bibpunct{(}{)}{;}{a}{}{,} 
\graphicspath{{figs/},{./}}

\newcommand{\revone}{}
\newcommand{\revtwo}{}
\newcommand{\revthree}{}

\newcommand{\ujlisti}{
\itemsep=0em
\parsep=0.5em
\partopsep=0.25em
\topsep=0em}

\def\degdot{{}^{\circ}\hspace{-0.3em}.}
\newcommand{\sgn}{{\rm sgn\,}}

\begin{document}

\title{Solar active region scaling laws revisited}

\author{Guilherme A. L. Nogueira\inst{1}
        \and Robertus Erd\'elyi\inst{1,2,3}
        \and Ruihui Wang\inst{4,5}
        \and Kristof Petrovay\inst{1}}

\institute{$^1$Dept.~of Astronomy, Institute of Physics and Astronomy,
ELTE E\"otv\"os Lor\'and University, Budapest, Hungary\\
$^2${Gyula Bay Zolt\'{a}n Solar Observatory (GSO), 
Hungarian Solar Physics Foundation (HSPF), Gyula, Hungary}\\
$^3$Solar Physics and Space Plasma Research Centre, 
School of Mathematical and Physical Sciences, 
University of Sheffield, UK\\
$^4$School of Space and Earth Sciences, Beihang University, Beijing, 
P.~R.~China\\
$^5$Key Laboratory of Space Environment Monitoring and Information
Processing of MIIT, Beijing, P.~R.~China\\
 \email{k.petrovay@astro.elte.hu}
}

\abstract
{The systematic variation of solar active region (AR) properties with their
magnetic flux has been the subject of
numerous studies but the proposed scaling laws still vary rather
widely.}
{A correct representation of these laws and the deviations from them 
is important for modelling the source
term in surface flux transport and dynamo models of space climate
variation, and it may also help constrain the subsurface origin of
active regions.}
{Here we determine active region scaling laws based on
the recently constructed ARISE active region data base listing bipolar
ARs for cycle 23, 24 and 25.} 
{For the area $A$, pole separation
$d$ and tilt angle $\gamma$ we find the following scalings against
magnetic flux $\Phi$ and heliographic latitude $\lambda$: 
$A \propto \Phi^{0.84}$, $\langle d\rangle =
3{\degdot}32\,\log(\Phi/\Phi_0)$, and $\langle\gamma\rangle \simeq
28{\degdot}62\,\sin\lambda$, with
$\Phi_0=1.6\times10^{20}\,\mathrm{Mx}$.
Residuals from these relations are also modelled.}
{These scaling relations are recommended for use in space climate
research for the modelling of future data or missing past data, as
well as for the identification of candidate rogue ARs.}


\maketitle
\nolinenumbers

\section{Introduction} Solar active regions (ARs) are strongly
magnetized regions of the solar atmosphere, formed by the emergence of
magnetic flux bundles from the subsurface layers
(\citealt{vDG+Green:LRSP, Fan:LRSP}). Their size can vary in a broad
range. From a physical point of view the relevant parameter
characterizing their size is the magnetic flux contained in the
emerging flux tube. As each emerged field line intersects the surface
twice, this flux can be approximated as half the total amount of
unsigned magnetic flux integrated over the photospheric area of the
AR. This flux measure $\Phi$ will generally depend on time (due to
flux emergence and cancellation across the neutral line) and on the
definition of the AR boundary.

The distribution of AR flux over the solar surface may vary. Hence,
other measures of AR size such as sunspot area or plage area may not
scale completely linearly with $\Phi$. In addition, the emergence
process typically results in a bipolar structure consisting of two
opposite polarity flux patches, of mean area $A$. The line connecting
the weighted mean positions of these patches is characterized by its
length, the pole separation $d$, and by its tilt angle $\gamma$ to the
azimuthal direction {\revtwo of the heliographic coordinate frame}.

How different AR characteristics, in particular $A$, $d$ and $\gamma$,
scale with $\Phi$ has been the subject of numerous studies. Due to the
large intrinsic scatter in AR properties very large samples 
{\revtwo containing thousands of ARs} 
are needed to draw statistically robust conclusions. A further complicating
factor is that all parameters vary with time during the evolution of
an AR, and further parameters such as heliographic latitude $\lambda$
or solar cycle phase and amplitude may also come into play. As a
result, despite numerous studies, firm quantitative conclusions
regarding the form of the scaling laws are still not available
{\revtwo (e.g., 
\citealt{Sheeley1966,Wang1989,Fisher1995,Meunier2003,Tian2003,Lemerle2015,vDG+Green:LRSP}).
}

Nevertheless, determining or at least constraining the scaling laws
would be important for several reasons. Firstly, these relations hold
important clues for the originating depth and emergence mechanism of
the subsurface magnetic flux tubes giving rise to ARs
(\citealt{Fan:LRSP}). Second, surface flux transport models widely
used to compute the evolution of the Sun's large scale magnetic
field include ARs as a source term (\citealt{Yeates2023}). This source
often needs to be modelled to account for missing data or future
evolution: it is clearly important for any such model to be as
realistic as possible. Finally, as it has been realised that the Sun's
axial dipole moment at the end of a solar activity cycle is a good
precursor of the amplitude of the next cycle (\citealt{Petrovay:LRSP})
and this dipole moment results from the summed contributions of
individual ARs, AR scaling laws allowing the calculation of these
contributions for a given distribution of ARs in time, latitude and
flux have an important role in solar cycle prediction.

One widely accepted assumption concerning the scaling laws, supported
(or at least not contradicted) by observational data is that the tilt
angle $\gamma$ is primarily determined by heliographic latitude
$\lambda$ (Joy's law), while the area $A$ and the pole separation
$d$ scale with the size of active regions $\Phi$. All these
relationships are increasing 
{\revtwo functions}, 
but regarding their form (e.g. value of
exponent, if modelled as a power law) there is considerable
disagreement.

{\revtwo
Without consulting actual data, a plausible first guess for the scaling 
relationships would be in the form of power laws}
\begin{equation}
A = C_A \Phi^k     \qquad  d = C_d \Phi^m  
\qquad \gamma = C_\gamma (\sin\lambda)^n
\label{eq:scalings}
\end{equation}
where $C_A$, $C_d$ and $C_\gamma$ are constants. 
{\revtwo For the first two exponents, the values $k=1$, $m=0.5$ are 
plausible first choices, reflecting a simplified scenario where bipolar
ARs are represented by two circular flux patches of fixed field strength,
tangential to each other. Assuming that the tilt of the AR axis is related to
the Coriolis force suggests the choice $n=1$.}
Detailed analyses based on observational data, however,
often yield rather different values and even the form of the suggested
scaling laws is in doubt in some cases. (See detailed discussions with
references below, in Section~3.)

Motivated by the newly acquired importance of AR scaling laws for space
climate prediction
{\revtwo  (\citealt{Bhowmik2023}),}
in this paper we make a new attempt at
constraining the scaling laws based on a recently constructed AR
database. Section~2 presents this input data set and its
pre-processing to derive the AR parameters under study. Section~3
presents the resulting scaling laws and the distribution of the
residuals from these laws. Section~4 discusses the implications of these
findings; finally, Section~5 concludes the paper.

\section{Data {\revtwo processing}}

\subsection{Sample}

Our input data were taken from the recently constructed ARISE
database\footnote{\tt
\url{https://github.com/Wang-Ruihui/A-live-homogeneous-database-of-solar-active-regions}} 
of solar active regions.  ARISE is a recently constructed database containing
basic parameters of bipolar solar active regions, extracted from SOHO/MDI and
SDO/HMI synoptic magnetic maps. 
{\revone 
During the construction of ARISE, ARs were
detected based on morphological operations and region growing and their
properties were extracted automatically by an algorithm, applying a bipolarity
condition and a size threshold. 
{\revtwo 
Specifically, in our detection algorithm, a region-growing module is applied to
determine the boundary of each active region. A magnetic field threshold of 50 G
for MDI and 30 G for HMI is used during the region-growing process. The
different thresholds are adopted to maintain consistency between the MDI and HMI
detection results. In addition, an area threshold of 351 pixels ($\simeq
412\,$Mm$^2$) is
imposed to remove small regions.
}
For further details the reader should consult the
publication describing the database (\citealt{ARISE1}).

For each AR, parameters of the Northern (positive) and Southern (negative)
magnetic polarity parts (here denoted by subscripts N and S) are listed
separately in the catalogue.\footnote{{\revone Note for readers less familiar with
magnetic data: Northern and Southern polarity here refer to the
directionality of magnetic field lines and they are not related to heliographic
latitude.}} These listed parameters include  heliographic latitude ($\lambda_N$,
$\lambda_S$), longitude ($\phi_N$, $\phi_S$), area ($A_N$ $A_S$) and
magnetic flux ($\Phi_N$ and $\Phi_S$).  
}

The database is regularly updated, i.e.
it is a "living" database. The data used for this study are the
version where recurrent ARs have been removed (\citealt{ARISE2}). This covers the
period from Carrington Rotation (CR) 1909 to 2290 (May 1996 - October
2024), corresponding to Solar Cycles 23, 24 and the first half of
cycle 25. The total sample is composed of 3005 bipolar ARs.
{\revtwo 
While some earlier studies (e.g., \citealt{Wang1989,Lemerle2015}) were based on
samples of comparable size, our input data are based on the more precise HMI and MDI
measurements. Our study, however, mainly differs from previous work in the
methods of data analysis, as discussed below.
}

\subsection{Quantities under study}

From the data listed in the database for each AR we analyze the
relationships between the following quantities.

\begin{itemize}

\item[$\Phi$:] Total absolute magnetic flux, calculated as
$\Phi=(|\Phi_N|+|\Phi_S|)/2$ where $\Phi_N$  and $\Phi_S$ are the
magnetic fluxes in the northern and southern polarity parts of the AR,
respectively. $\Phi$ is given in units of maxwell [Mx], or, in some
cases SFU (solar flux unit, \citealt{Sheeley1966}). 1\ SFU$=10^{21}\,$Mx.

\item[$A$:] Total area per polarity, calculated as
$A=(A_N+A_S)/2$ where $A_N$  and $A_S$ are the
areas of the northern and southern polarity parts of the AR,
respectively. $A$ is given in units of microhemisphere [MSH].

\item[$\lambda$:] Heliographic latitude $\lambda=(\lambda_N+\lambda_S)/2$, given in degrees.

\item[$d$:] Pole separation, i.e. the distance between the centers of the
N and S polarity parts, calculated from the spherical cosine theorem
\begin{equation}
\cos d = \sin\lambda_N \sin\lambda_S + \cos\lambda_N \cos\lambda_S
\cos(\phi_N-\phi_S)
\end{equation}
where $\phi$ and $\lambda$ are heliographic longitude
and latitude, respectively. $d$ is given in [heliocentric] degrees.

\item[$\gamma$:] Tilt angle, {\revthree expressed in degrees} 
and defined as
\begin{equation}
\gamma=\arctan\frac{\lambda_S-\lambda_N}{\cos\lambda\, (\phi_N-\phi_S)}
\end{equation}
Note that this formula is a Euclidean approximation to the azimuth of the
direction of the trailing (lower heliographic longitude) polarity
from the vantage point of the leading (higher longitude) polarity, the
azimuth being measured northwards from east. The formula does not
distinguish ARs following Hale's polarity rules from those opposing it
(non-Hale ARs).
{\revthree Scaling laws for the non-Hale regions, constituting
$\sim 5$\% of all ARs (\citealt{Munozjara2021}), 
will be the subject of a follow-up study.}

As the typical sign of $\gamma$ is opposite on the two hemispheres
in accordance with Joy's law, we also introduce the alternative form
\begin{equation}
\gamma_J=\gamma\cdot\sgn\lambda
\end{equation}
which has the same sign on both hemispheres for ARs adhering to Joy's
law. Thus,
$\langle\gamma_J\rangle$ measures overall adherence to Joy's
law in a given population; 
$\langle|\gamma|\rangle$ characterizes the preferred azimuthal
orientation (East--West vs.\ North--South);
while $\langle\gamma\rangle$ typifies hemispheric asymmetry.

\end{itemize}

\begin{figure}[htb]

\includegraphics[width=\columnwidth]{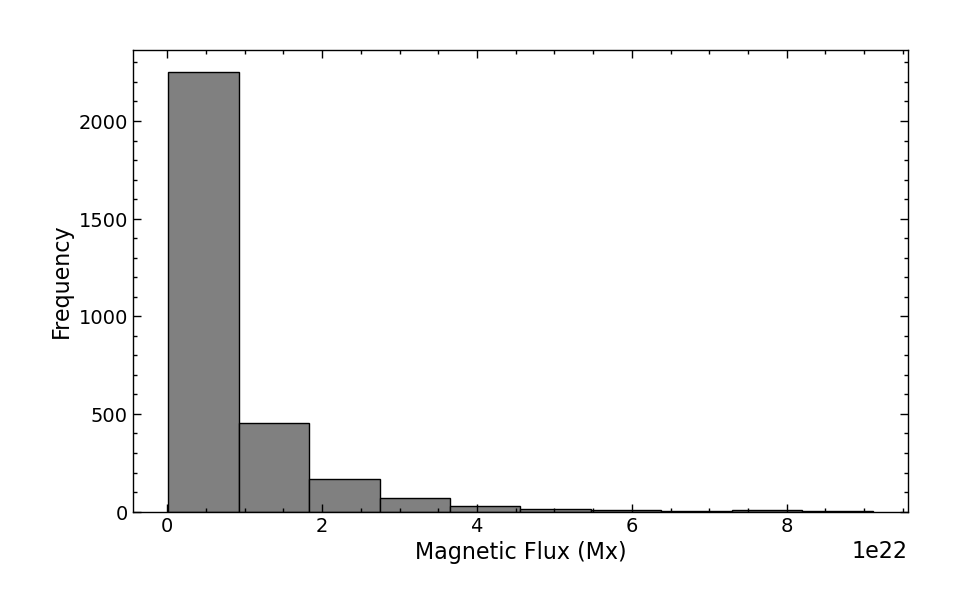}
\includegraphics[width=\columnwidth]{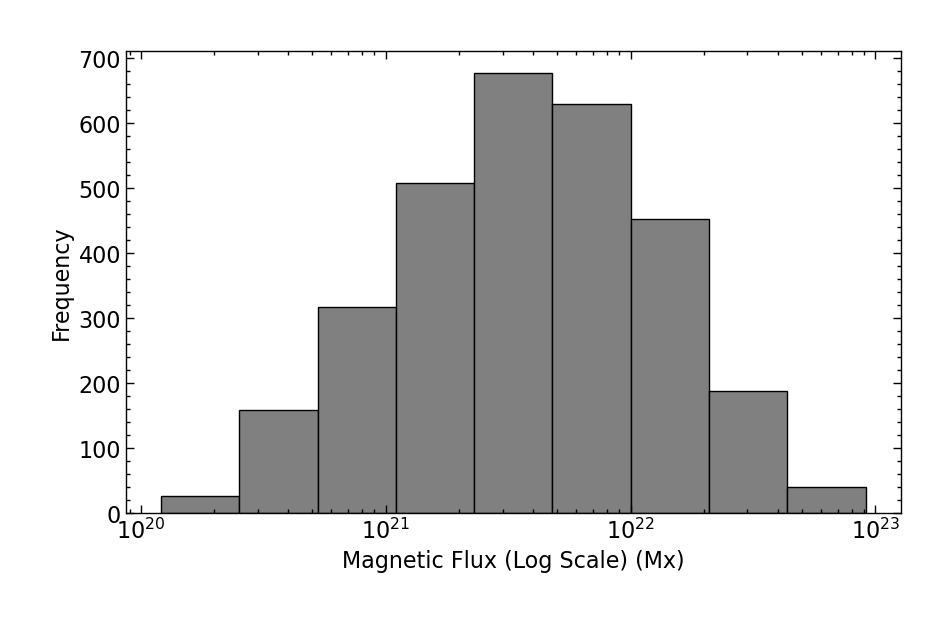}
\caption{Histogram of the total flux $\Phi$ of active regions on
linear (left) and logarithmic (right) scale}
\label{fig:phihist}
\end{figure}

\begin{figure}[htb]

\includegraphics[width=\columnwidth]{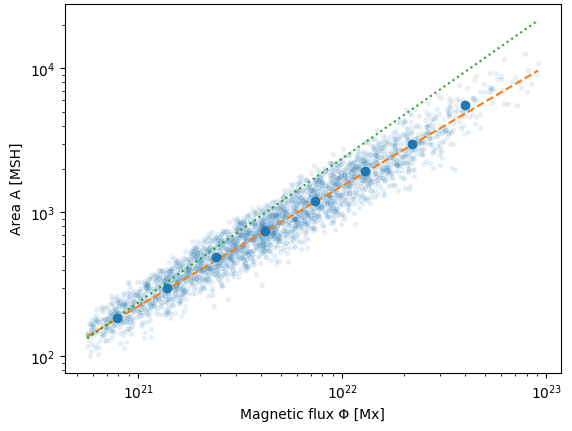}
\caption{Plot of active region area $A$ vs.\  magnetic flux $\Phi$,
with power law fits to the medians of the binned data 
(blue circles with error bars).
The dashed line corresponds to the optimal fit;
the dotted line shows a linear relationship $A\sim\Phi$ for comparison.
Lighter background is a scatterplot of the individual points.}
\label{fig:A_Phi}
\end{figure}

\begin{figure}[htb]

\includegraphics[width=\columnwidth]{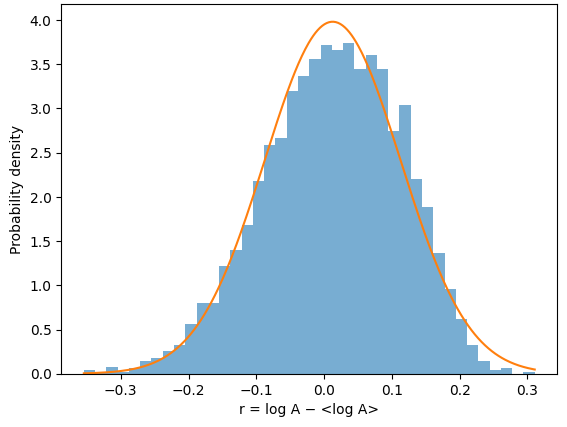}
\caption{Histogram of the residuals from the  fit in
Fig.~\ref{fig:A_Phi}. The solid line is a Gaussian fit.}
\label{fig:A_Phi_residual_histog}
\end{figure}

\begin{figure}[htb]

\includegraphics[width=\columnwidth]{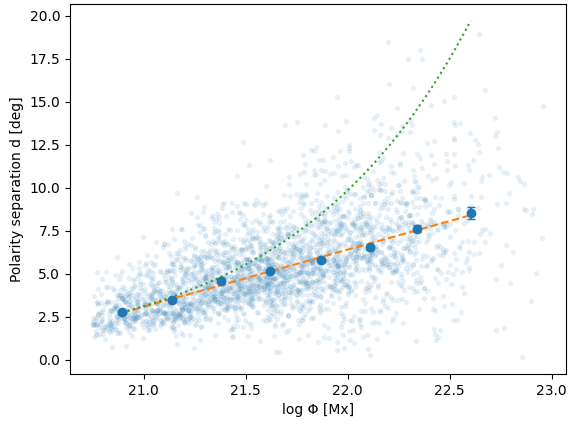}
\caption{Plot of the polarity 
separation $d$ vs.\ logarithm of active region flux $\Phi$. 
Median values are plotted for each bin (blue circles with error bars).
The dashed line is a linear (i.e.\ logarithmic) fit. The dotted line shows
the scaling $d\sim\Phi^{1/2}$ for comparison.
Lighter background is a scatterplot of the individual points.}
\label{fig:d_Phi}
\end{figure}

\begin{figure}[htb]

\includegraphics[width=\columnwidth]{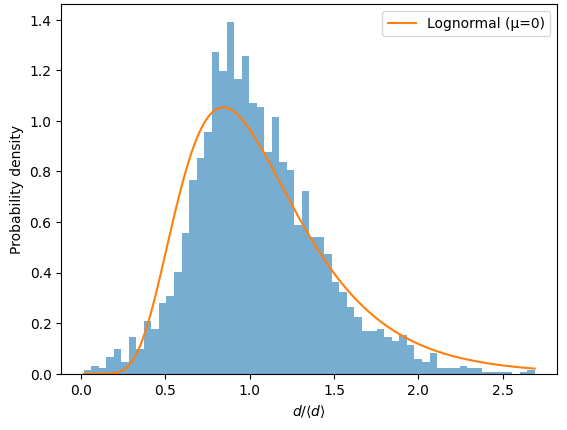}
\caption{Histogram of the fractional residuals $d/\langle d\rangle$
relative to the linear fit in Fig.~\ref{fig:d_Phi}. 
The solid curve is a lognormal fit.}
\label{fig:d_Phi_residfract_histog}
\end{figure}

\subsection{Cycle assignment}

In a first approximation, cycle assignment was based on the date of
the observation, using the official starting date of each cycle as
given in the SIDC/SILSO database.\footnote{\tt
\url{https://www.sidc.be/SILSO/cyclesminmax}} As, however, cycles are
known to overlap by up to 2 years, this first cycle assignment was
manually corrected for high latitude ARs in the last 20 CRs and for
low latitude ARs in the first 20 CRs during each cycle. The dividing
line between "high" and "low" here lies roughly at
$|\lambda|=15^\circ$ latitude but as the latitudinal distribution in
these time periods displays a very clear bimodality, there was no
ambivalence in selecting the ill-assigned ARs.  The corrected cycle
attributions were also compared with the assignments given by
\cite{Leussu2017} wherever possible.

This resulted in manual correction of the cycle assignment for 30 ARs.

\section{Results}

Histograms of the flux values and their logarithms are shown in
Fig.~\ref{fig:phihist}.  These histograms are subject to a heavy bias
for lower fluxes due to the field strength threshold (50 G/30 G in
MDI/HMI data), area threshold (412 Mm$^2\simeq 4.2\,$MSH) and
morphological transformations applied during the compilation of the
ARISE database. Nevertheless, irrespective of the effects shaping the
distribution, it can be seen that the histogram of the flux values is
strongly skewed, with a long tail. The histogram of $\log\Phi$, in
contrast, is much less skewed. 
{\revone This is in agreement with the known result that the size distribution
of larger ARs, less affected by selection effects, is approximately lognormal
(\citealt{Harvey1993}, \citealt{Seiden1996}, \citealt{Munozjara2015}).}
Hence we opt to group our data into
bins equidistant in $\log\Phi$. Nine bins are introduced; the bins
have widths of 0.25, except the lowest and highest bins that comprise
all ARs with  $\log\Phi< 20.75$ and  $\log\Phi> 22.5$, respectively.
The lowest flux bin, most heavily influenced by the threshold effects,
is discarded.

For variables with a non-normal distribution the use of the median is
expected to be more robust than using the mean. Hence, for each bin,
the median value of the variable studied is calculated. The
uncertainty of this value is estimated as $\sigma_i/\sqrt{n_i}$,
with $n_i$ the number of values in the bin and $\sigma_i$ their
standard deviations. While this formula is strictly only valid for a
normal distribution, the error introduced by this is considered
admissible in view of the large extra computational burden that a
proper bootstrap estimate would imply.
{\revone 
This approximation nevertheless calls for due reservation when
evaluating  fits with a significantly non-Gaussian scatter and resulting in
$p$-values that are not close to either 1 or 0.
}

\subsection{Area vs.\ flux}

For the relation between flux and area, a log--log representation of
the results is shown in  Fig.~\ref{fig:A_Phi}. A linear regression
(dashed) provides a very good representation of the data
($\chi^2=1.83$, $p$-value 0.78):
{\revone
\begin{equation}
\log A = k \log\Phi + \log C_A
\label{eq:A_Phi}
\end{equation}
with 
$k=0.836{\pm}0.005$ and $\log C_A=-15.21{\pm}0.11$. 
(Recall that $\Phi$ is given in units of Mx and $A$ is given in MSH.)
The fit corresponds to a power law dependence of the form
$A = C_A \Phi^k$ with $C_A=6.17\cdot 10^{-16}$, using the same units. The case
$k=1$ (dotted) is clearly excluded.
}

The histogram of logarithmic residuals from equation~\ref{eq:A_Phi} is
displayed in Fig.~\ref{fig:A_Phi_residual_histog}. Despite a distinct
negative skew (skewness -0.28), a Gaussian fit with standard deviation
0.1 provides a reasonable representation of the data. This means that
for a bipolar region of magnetic flux $\Phi$ the logarithm of the area
of the individual polarity patches is best represented as $\log
A=\langle\log A\rangle + r_A $ where $\langle\log A\rangle= k\log
\Phi+\log C_A$ and $r_A$ is a Gaussian random
variable with standard deviation $0.1$.

The analysis was repeated separating individual solar cycles: all cycles
were found to follow the scaling (\ref{eq:A_Phi}), without statistically
significant deviations.

{\revone The power-law scaling found here is in good agreement with the
findings of \cite{Meunier2003} who reported $\Phi\sim A^{1/k}$ with $1/k\simeq
1.2$ for the inverse relation. Other}
previous studies of the relationship between magnetic flux and area in
solar active regions all resulted in linear scalings, i.e.\ $k=1$
(\citealt{Sheeley1966,Wang1989,vDG+Green:LRSP,Murakozy2024}). These
studies, however, were mostly limited to sunspots; they did not
normally present a statistical test of the goodness of the fits
suggested; and the flux estimates used in early studies were
{\revtwo affected by large errors.}

It follows from our result that the average field strength $\Phi/A\sim
\Phi^{0.16}$. While this is a mild increase, over the two orders of
magnitude in flux covered by our sample ($10^{21}$--$10^{23}$Mx)  it
still implies a factor of two increase in the average field amplitude.
Whether this increase is due to a higher fraction of the area covered
by spots or to a higher overall plage field strength is unclear.
Further research should shed light on this issue.

\subsection{Pole separation vs.\ flux}

Fig.~\ref{fig:d_Phi} plots $d$ against $\Phi$ for all ARs in the data base. 
It is apparent that a logarithmic fit of the form
\begin{equation}
\langle d\rangle = m_d \log(\Phi/\Phi_0) 
\label{eq:d_Phi}
\end{equation}
provides a very good description of the data ($\chi^2=0.8$, $p$-value
0.997). 
The best fit parameters obtained by linear regression are $m_d=3.32{\pm}0.11$ and 
$\Phi_0=0.16\,$SFU. {\revtwo (Uncertainty in $\Phi_0$ is indirectly determined
by the uncertainty of the intercept as $m_d \log\Phi_0 = -2.64{\pm}0.05$.)}

The relation between $\Phi$ and $d$ has been considered in a
surprisingly low number of studies. \cite{Wang1989}, \citet{Tian2003}
and \cite{Lemerle2015} all reported a power-law relationship, i.e.
$d\sim\Phi^m$, however the exponents obtained varied
substantially, values being $m=0.77$, $m=0.87$ and $m=0.42$,
respectively.  This was based on a direct linear regression fit to an
unbinned log--log scatterplot in all cases, and the goodness of the
fit was not determined. The high $p$-value obtained in our analysis
clearly shows that a logarithmic fit is superior to the power law
fits. We note that a first indication of the logarithmic relationship
can also be seen in Fig.~4 of \cite{Erofeev2023} where, based on
white-light images, the plotted relation between the logarithm of the
total sunspot area in a sunspot group vs.\ the pole separation of
clearly bipolar groups is not too far from linear.

Fig.~\ref{fig:d_Phi_residfract_histog} presents the histogram of the
fractional residuals $r_d=d/\langle d\rangle$ relative to the
mean law given by eq.~\ref{eq:d_Phi}. A lognormal fit is found to be a
reasonably good representation of the data. 
Accordingly, for a bipolar region of magnetic  flux $\Phi$ the pole
separation may be best represented as $d=r_d\cdot\langle d\rangle$ where
$\langle d\rangle= m_d \log(\Phi/\Phi_0)$ as above and $\log(r_d)$ is
a normally distributed random variable, with standard deviation 0.41.

\begin{table}[htb]

\caption{Bulk characteristics of the tilt in the sample}
\label{table:bulktilts}

\begin{tabular}{lrrr}
&&&\\
\hline
&&&\\
& \multicolumn{3}{c}{{\revone Tilt values}  $[^\circ]$} \\
&&&\\
 & $\langle\gamma\rangle$ & $\langle\gamma_J\rangle$ 
 & $\langle|\gamma|\rangle$  \\
&&&\\
\hline
&&&\\
\ median	     & -0.50 &    7.95&    14.09  \\
\ mean  	     & -0.62 &    6.57&    18.48  \\
\ st.\ dev. $\sigma$ & 24.94 &   24.07&    16.76  \\ 
\ error $\sigma/\sqrt{N}$  & 0.45 & 0.44  & 0.31 \\
&&&\\
\hline
\end{tabular}

\end{table}

\begin{figure}[htb]

\includegraphics[width=\columnwidth]{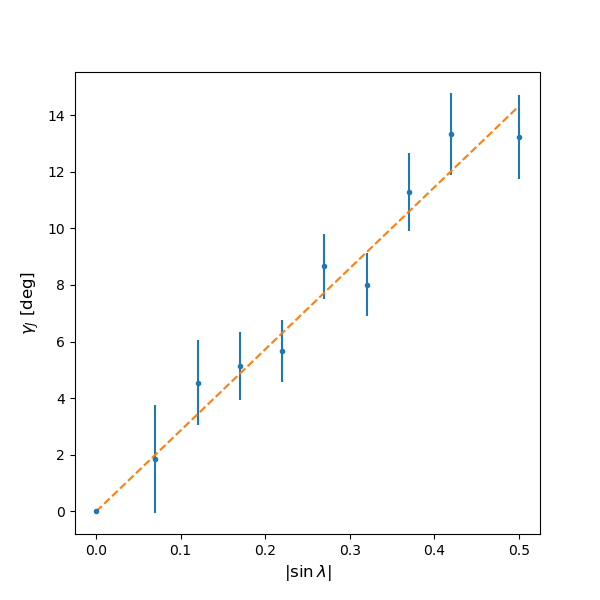}
\caption{Plot of the tilt angle following Joy's convention, 
$\gamma_J$ vs.\ sine latitude for the complete sample.  The
dashed line is a homogeneous linear fit. }
\label{fig:stats_joy}
\end{figure}

\begin{figure}[htb]

\includegraphics[width=\columnwidth]{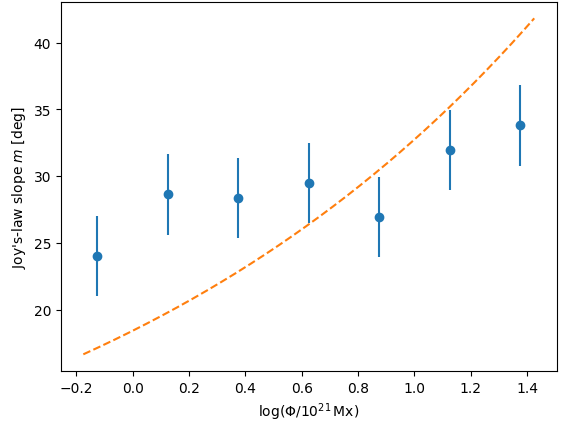}
\caption{Slope of Joy's law determined for subsamples in different flux
bins plotted against $\log\Phi$. The dashed line is an optimal fit of
the form $m_J\sim\Phi^{1/4}$, clearly inconsistent with the data.}
\label{fig:joyslope_Phi}
\end{figure}

\subsection{Tilt vs.\ latitude (Joy's law)}

Bulk characteristics of the tilt distribution are summarized in
Table~\ref{table:bulktilts}.  The complete sample shows a clear
preference for a tilt of $\gamma_J\sim 7^\circ$, with the leading pole
positioned closer to the equator, in accordance with Joy's law. The
numbers in the table align well with the findings of \cite{Qin2025}
and they place the ARISE sample among the higher-tilt ones, as
discussed by \cite{Erofeev2023}. This is in line (\citealt{Wang2015})
with 
{\revtwo
the database used here being restrcted to larger ARs
}
with a clearly bipolar structure and good polarity balance. Note that
discrepancies between data sets may also be related to the presence of 
magnetic tongues in plage structure (\citealt{Poisson2020}). 

Hemispheric asymmetry, as characterized by  $\langle\gamma\rangle$, is 
not significant. 

In contrast to area and pole separation, the tilt angle has been known
to be primarily determined by heliographic latitude rather than flux
(Joy's law). To study this relation, we choose $\sin\lambda$ as
independent variable. This is motivated by the consideration that
Joy's law most plausibly originates from the Coriolis force which
scales with $\sin\lambda$. Nine bins are introduced; the bins have
widths of 0.05, except the lowest and highest bins that comprise all
ARs with $|\sin\lambda| < 0.1$ and $|\sin\lambda| > 0.45$,
respectively.

Figure~\ref{fig:stats_joy} presents the binned data with the Southern
hemisphere folded over the Northern one, {\revone 
i.e. as a function of $|\sin\lambda|$.}
A simple one-parameter linear 
{\revtwo regression}
of the form
\begin{equation}
\langle\gamma_J\rangle= m_J \sin\lambda   \qquad m_J=28{\degdot}62{\pm}1{\degdot}44
\label{eq:joylaw}
\end{equation}
is clearly a perfectly good representation of the data ($\chi^2=0.54$,
$p$-value 0.9998).  We are thus unable to confirm occasional claims in
the literature (\citealt{McClintock2013,Erofeev2023}) of various
nonlinearities in the shape of Joy's law.

\begin{figure}[htb]

\includegraphics[width=\columnwidth]{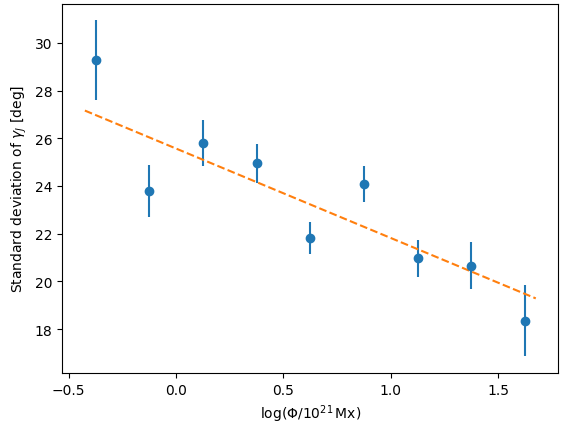}
\includegraphics[width=\columnwidth]{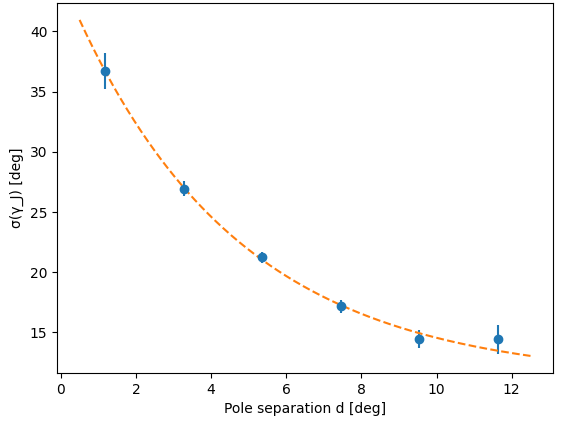}
\caption{Plot of the standard deviation of tilt angles $\gamma_J$
vs. magnetic flux (top) and pole separation (bottom). 
The dashed lines are linear end exponential fits, respectively.}
\label{fig:joyscatter}
\end{figure}

\begin{figure}[htb]

\includegraphics[width=\columnwidth]{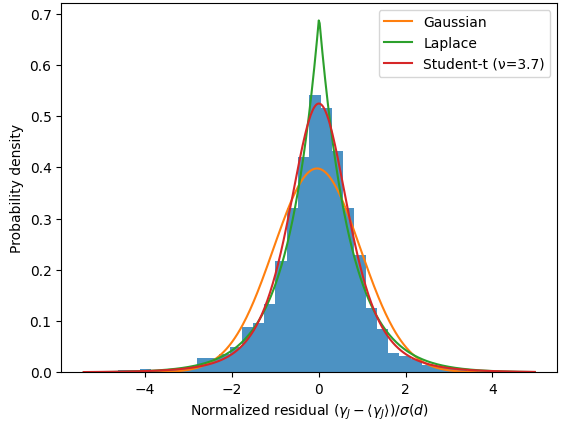}
\caption{Histogram of normalized residuals around Joy's law. Optimally
fitted Gaussian, Student's $t$ and Laplace distributions are shown for
comparison.}
\label{fig:joyslaw_residuals_dnorm}
\end{figure}

Separating the data by hemisphere or by solar cycle, no statistically
significant differences in the value of the coefficient in Joy's law
were found. Our results for cycles 23 and 24 agree with the findings of
\cite{Will2024} in that the value of $m_J$ is higher in cycle 23 but
the difference is not significant. The form of our fitting function,
forced to go through the origin may have a role in the lack of
hemispheric asymmetry in our study. As recently pointed out by
\cite{Zeng2024}, the hemispheric asymmetry apparent in some studies is
mostly due to low-latitude regions, and this would only show up when a
non-zero intercept is allowed for.

While the primary determinant of the tilt is clearly the latitude,
some theoretical and empirical studies suggest that magnetic flux or
AR size may have a secondary role (\citealt{Howard1993, Fan1994,
Fisher1995, Sreedevi2024, Qin2025}). In Fig.~\ref{fig:joyslope_Phi} we
plot the value of the slope $m_J$ for subsamples divided into flux
bins, as used in the previous sections. (The top and bottom flux bins
are not shown as here some latitude bins contain too few points for
reliable fits.) The plot is strongly suggestive of an increasing
trend, but the null hypothesis of no dependence cannot be discarded
with more than $\sim 1\sigma$ confidence, while the classic
theoretical prediction $m_J\sim\Phi^{1/4}$ (\citealt{Fan1994}) is
clearly inconsistent with the data. If only three flux bins are used
(with divisions at 2 and 7 SFU), the resulting Joy slopes in order of
increasing flux are $m_J= 26{\degdot}1{\pm}3{\degdot}0$,
$28{\degdot}4{\pm}2{\degdot}1$ and
$33{\degdot}1{\pm}2{\degdot}2$, respectively. The increasing trend is again
suggestive, while the null hypothesis cannot be discarded at a
confidence level much better than  $1\sigma$. (This also illustrates
the low sensitivity of our findings to the choice of binning, not
demonstrated here in each and every case.)

These inconclusive results {\revtwo explain} 
contradictory results in previous
studies, some of which reported no correlation or even negative
correlations of tilt with flux (\citealt{Kosovichev2008,
McClintock2016, Jha2020}).

The standard deviation of the tilt values, on the other hand, displays
a much clearer trend with magnetic flux (Fig.~\ref{fig:joyscatter}),
significant at $6\sigma$. A linear best fit results in
\begin{eqnarray}
&&\sigma_J =a+b\log(\Phi/1 \mbox{SFU}) \quad \mbox{with}
\label{eq:joyscatter_Phi} \\
  &&a = 25{\degdot}57{\pm}0{\degdot}51^\circ \quad \mbox{and} \quad 
  b =  3{\degdot}75{\pm}0{\degdot}61^{\circ} \nonumber
\end{eqnarray}

Given our above finding $d\sim\log{\Phi}$, we also construct a plot of
$\sigma_J$ vs.\ $d$ (Fig.~\ref{fig:joyscatter}, right panel). The
alignment of the points becomes even tighter in this case, 
{\revtwo apparently}
indicating that the geometry of the rising loop plays a more direct role in
regulating deviations from Joy's law. An exponential fit is found to
be a {\revone convincing description ($p$-value 0.73)} of the relation:
\begin{eqnarray}
&&\sigma_J =\sigma_\infty + A\exp (-d/d_0) \qquad \mbox{with}
\label{eq:joyscatter} \\
  &&\sigma_\infty = 11.0{\pm}1.7 \qquad A =
  33{\degdot}5{\pm}1{\degdot}9^\circ \qquad 
  d_0 =  4{\degdot}4{\pm}0{\degdot}7^{\circ} \nonumber
\end{eqnarray}
Note, 
{\revtwo however,} 
that the stronger dependence of the scatter on $d$ may at least
partly be caused by a simple geometrical effect. For an uncertainty
$\sigma_p$ in the determination of the pole position, the error in
tilt determination will be $\sigma_p/d$, decreasing with separation. 
{\revtwo 
This effect may also contribute to the tighter alignment of the points in the
lower panel.
}

{\revone Note also that the possibility that the poorer alignment of the scatter
against flux may be due to lower precision of the flux determination cannot be
completely discarded. However, we do not consider this possibility highly
likely, given that flux values in the ARISE catalogue were carefully
cross-checked comparing MDI vs HMI data and against other catalogues
(\citealt{ARISE1}).  Furthermore, \cite{Fisher1995} report a similarly tight
relation between $d$ and tilt scatter even though their $d$ values were
determined from white light images only, implying significantly larger
uncertainties. Finally, we note that \cite{Stenflo2012} report a much stronger
dependence of tilt scatter on $\Phi$ but only for flux values in the range
$0.1$--$1\,$SNU, below that studied here.}

It remains to determine the form of the distribution of the residuals
around Joy's law. The decreasing trend of scatter around the law,
equation~(\ref{eq:joyscatter}), suggests to use normalized residuals
$r_J=(\gamma_J- \langle\gamma_J\rangle)/\sigma_J$. The histogram of the
$r_J$ values using equation~(\ref{eq:joyscatter}) for $\sigma_J$ is
displayed in Fig.~\ref{fig:joyslaw_residuals_dnorm}. Clearly, with
this normalization, the full sample collapses onto a nearly universal
distribution. 

{\revone To model this distribution, we first attempt to fit a Gaussian, which
would be the natural expectation for tilt scatter resulting from random
convective buffeting during flux emergence. Many such independent convective
kicks should naturally result in a Gaussian distribution by virtue of the
central limit theorem of probability theory. The Gaussian fit, however, can be
rejected with a very high confidence ($p$-value of $3\cdot 10^{-8}$ based on a
Kolmogorov--Smirnov test). Even by visual inspection, the distribution  is
distinctly non-Gaussian and leptokurtic, with a strong peak and extended tails 
(excess kurtosis $+2.8$). The high excess kurtosis is the hallmark of an
intermittent stochastic perturbing process where most instances are only
slightly perturbed (the strong peak), while in a small fraction of instances a
single large perturbation results in large deviations (the tail). Standard
methods in statistical physics to model such leptokurtic distributions
resulting from intermittent processes include a double exponential
(a.k.a.~Laplace) distribution or Student's $t$-distribution. The Laplace
distribution fits our data only marginally ($p$-value 0.067), while a
satisfactory overall match is provided by Student's $t$-distribution with 3.7
degrees of freedom ($p$-value 0.18). }
This agrees with the findings of \cite{Munozjara2021} that a $t$-distribution
with $3.45$ degrees of freedom represents well the (unnormalized) tilt
residuals. 

The presence of extended tails thus indicates that fluctuations are governed by
intermittent, impulsive perturbations to the emerging flux tubes, rather than
many small, independent Gaussian kicks. This type of distribution may be
expected from sporadic convective buffeting, pre-existing magnetic structures,
or episodic vortical forcing during emergence, {\revtwo as a result of which a
small fraction of AR flux loops may be subjected to an intermittent
large-amplitude torque during their rise.}

The actual histogram of normalized residuals, however, has a
considerable skewness of $-0.4$. From visual inspection of
Fig.~\ref{fig:joyslaw_residuals_dnorm} this asymmetry is
primarily located in the tails, where negative residuals are more
common than the fitted $t$-distribution, while positive residuals are
less common. This seems to suggest that AR flux tubes subjected to
excessive disturbance prior to emergence tend to become completely
oblivious of Joy's law, their tilt distribution becoming more
symmetrical to the equator.

In summary, we suggest that for a bipolar region of magnetic flux
$\Phi$ emerging at latitude $\lambda$ the tilt angle $\gamma_J$ is
best represented as $\langle\gamma_J\rangle + r_J \sigma_J$ where
$r_J$ is a random variable with a distribution described by Student's
$t$-distribution of $3.7$ degrees of freedom, and $\sigma_J$ is
obtained from equation~(\ref{eq:joyscatter}) [or alternately from
(\ref{eq:joyscatter_Phi}), in which case a $t$-function with 2.9
degrees of freedom is to be used]. For applications
where an accurate representation of the tails of the distribution
matters, accounting for the skewness by suppressing [amplifying] the
positive [negative] tail of the distribution may be considered.

\section{Discussion}

\subsection{Evolutionary effects}

Active regions are evolving throughout their life time, implying that
all the studied quantities depend on time (\citealt{vDG+Green:LRSP},
\citealt{FDE2021}).
The AR data listed in the ARISE catalogue were obtained from synoptic
maps. On these maps, ARs are captured on the day of their central
meridian passage, i.e. at a random instant during their evolution. It is
therefore not necessarily trivial to link, e.g., the measured value of
the flux $\Phi$ to its maximal value $\Phi_0$, which presumably
corresponds to the physically meaningful total magnetic flux in the
rising magnetic flux loop. Other parameters like $d$ or $\gamma$ may
not display a maximum in their evolution but some characteristic
values corresponding to a certain evolution phase (e.g. when $\Phi$
attains its maximum) may still be defined. The question arises to what
extent the scaling relations derived above can be considered valid for
the more meaningful underlying parameters $\Phi_0$, $d_0$ etc.

Fortunately, there is some evidence that the parameter evolution
curves of ARs exhibit a certain universality, i.e., for any observable
$y$
\begin{equation}
y = y_0 f_y(t/T)
\end{equation}
where the AR lifetime $T=f(\Phi_0)$ and the function $f_y(x)$ is a
universal average AR evolution curve for the observable $y$. 
These average curves were recently determined by \cite{Svanda2025} for
different observables up to a time shortly after flux maximum for 36
ARs. The curves are generally consistent with findings from previous
studies with narrower focus or smaller samples
(\citealt{Kosovichev2008, Schunker2019, Will2024}).

Accepting the universality assumption, the expected value of an 
observable at a random instant, as measured on synoptic maps, will
scale linearly with the characteristic value $y_0$:  
\begin{equation}
\langle y\rangle = \frac 1T \int_0^T y_0 f_y(t/T)\,dt = y_0 \int_0^1
f_y(x)\, dx
\end{equation}
This suggests that the scalings found in this work may indeed be
valid also for the underlying characteristic scales, with a scatter
resulting from a combination of evolutionary phase and real physical
scatter.

Despite the evidence for a universal average behaviour, some doubts
regarding its validity linger, especially in the case of the tilt
(\citealt{McClintock2016}). Further, more extensive studies of AR evolution are
needed to clarify this issue. 
{\revone Studies like \cite{Schunker2020}, or}
the AutoTAB catalogue compiled by
\cite{Sreedevi:AutoTAB} are important steps in this direction.

\subsection{Implications for active region formation}

Models and concepts for the subsurface origin of active regions are
reviewed by \cite{Fan:LRSP} and \cite{Weber2023}. The classic paradigm
of the buoyant rise of a flux loop from the bottom of the convective
zone to near-surface layers is still the only coherent scenario
(\citealt{Petrovay_Christensen}) worked out in numerical detail.
Alternative possibilities include originating depths in the bulk of
the convection zone or in the near-surface shear layer,  {\revone and
rise driven by the drag of convective upflows (see, e.g., 
\citealt{Birch2016}, \citealt{Hotta2020} and\citealt{Chen2022}).}

Predictions of AR scaling laws from these models have focused on Joy's
law. Suggestions for the origin of this law include the following.

\begin{description}
\item{(1)} The tilt may reflect the orientation of the underlying flux
tubes giving rise to the emerging loops. The orientation of these
tubes, originating from the windup of the seed poloidal field present
at solar minimum, may be inclined to the azimuthal direction. 
(\citealt{Babcock1961, Norton2005, Tlatova2018})
\item{(2)} The tilt forms during the rise of the flux loop through the
convective zone due to the action of Coriolis force on
\begin{description}
\item{(a)} flows inside the loop whose rise is driven by magnetic
buoyancy (\citealt{DSilva1993, Caligari1995, Weber2013} etc.)
\item{(b)} helical convective flows distorting the loop (and possibly 
contributing to its rise) via the drag (``$\Sigma$-effect'', 
\citealt{Longcope1998})
\end{description}
\item{(3)} The tilt may form during and/or after the emergence of
the flux loop through the surface, due to the effect of supergranular
flows affected by the Coriolis force (\citealt{RolandBatty2025}, {\revone
\citealt{Schunker2025}}).
\end{description}

In the classic model of the origin of ARs, sometimes called the ``
buoyant thin flux tube paradigm'', the underlying toroidal field lies
in the tachocline, at or slightly below the bottom of the convective zone.
In such models several independent lines of evidence point to an
initial field strength of $B_0\sim 10^5$\,G (see
\citealt{Petrovay_Christensen} for a summary of these arguments). The
prediction by such models (\citealt{Fan1994}) for the tilt is
$\gamma_J\sim\Phi^{1/4}B_0^{-5/4}$. Finding evidence for this
dependence would be a decisive ``smoking gun'' in favour of the thin
flux tube model. However, as we have seen in Section 3.3, the observed
flux dependence of the tilt is much weaker than the predicted
$\Phi^{1/4}$ dependence. This is not necessarily in contradiction with
the classic model as the initial field strength may also vary, and
there may well be a statistical relation between $\Phi$ and $B_0$.
Indeed, as discussed in Section 3.1 above, our $A$--$\Phi$ relation
implies such a relation between the flux of an AR and its  mean
magnetic field in the photospheric layers.

{\revtwo 
Concerning the pole separation $d$,
}
observations of AR evolution generally indicate that after an initial
rapid increase, $d$ saturates at a constant value of $\sim 100$ Mm
(\citealt{Kosovichev2008, Schunker2019, Svanda2025}). Indeed,
recurrent sunspot groups are often seen to return several times with
the position of the polarities hardly changing at all. This is a
surprising fact in the light of the buoyant flux loop paradigm as the
most unstable modes tend to be those with low wavenumber $m$,
corresponding to scales of 300 Mm or longer. The observed length
scales lie closer to the scales of turbulent convection in the deep
convective zone in mixing-length models and numerical simulations,
determined by the scale height. It may then be that the typical scale
of finite amplitude initial perturbations plays a more important role
in determining the size of the rising flux loops. The puzzling but
very robust logarithmic scaling of $d$ with $\Phi$ discovered in our
analysis  may hold important clues regarding turbulence spectra in the
deep convective zone and originating depths of the rising flux loops.

\section{Conclusion}

In this paper we have analyzed the recently compiled ARISE database of
solar bipolar magnetic regions to study how {\revtwo geometric} AR parameters scale
with the fundamental parameter, the magnetic flux $\Phi$. The {\revtwo
geometric}
parameters studied were the area $A$, the pole separation $d$ and the
tilt angle $\gamma_J$.

Our most novel finding is that, contrary to what was found (or rather,
a priori assumed) in earlier studies the $d$--$\Phi$ relation is not a
power law but a well determined and highly robust logarithmic
relation. The scaling of $A$ with $\Phi$ is found to deviate from
linear, implying that the mean field strength increases with region
size. For the tilt angle we find that the slope of Joy's law shows a
tendency to increase with $\Phi$ but the significance of this result
is low and the trend is much weaker than the theoretical prediction
$\gamma_J\sim\Phi^{1/4}$. Scatter around Joy's law decreases linearly
with $\Phi$ and exponentially with $d$.

The distribution of residuals around the mean scaling laws was also
studied. $A$ and $d$ were found to be roughly lognormally distributed,
while we confirm the earlier finding that residuals from Joy's law
follow Student's $t$-distribution with $\sim 3.5$ degrees of freedom.
These scalings and residual distributions allow us to construct a
recipe for the synthesis of an ensemble or population of active
regions correctly reflecting the observed statistics. Collecting the
relevant fitting formulae from the main text of the paper, this recipe
is summarized in the Appendix.

\begin{acknowledgements}
This research was supported by the European Union's Horizon 2020
research and innovation programme under grant agreement No.~955620 and
by the NKFIH excellence grant TKP2021-NKTA-64. RE is also grateful to
the Hungarian National Research, Development and Innovation Fund
(NKFIH, grant no.~K142987);  the UK Science and Technology Facilities
Council (STFC, grant no.~ST/M000826/1); PIFI (China, grant
no.~2024PVA0043).
{\revtwo 
RHW is supported by the National Natural Science Foundation of China 
(grant No. 12425305).
}
\end{acknowledgements}

\bibliographystyle{aa}
\bibliography{ARscaling}

\begin{appendix}

\section{Active region population synthesis: a recipe}

Surface flux transport models, widely used to compute the evolution of
the Sun's large scale magnetic field, include ARs as a source term
(\citealt{Yeates2023}). This source often needs to be modelled to
account for missing data or future evolution: it is clearly important
for any such model to be as realistic as possible. 

In what follows we give a concise summary of the relevant findings in
our paper in the form of a ``recipe'' to generate an ensemble of
active regions to be used as source term in an SFT or dynamo model.
Prescribing when, where and with what magnetic flux a bipolar region
will emerge in such models is beyond the scope of the present work. We
restrict our attention to determining the area $A$ of the individual
flux patches; their separation $d$; and the tilt angle of the bipole
axis.

For a bipolar region of magnetic flux $\Phi$  emerging at latitude
$\lambda$:

(1) The logarithm of the area $A$
of the individual polarity patches is best represented as 
$\log A=\langle\log A\rangle + r_A $
where 
\begin{equation}
\langle\log A \mbox{[MSH]}\rangle= k\log\Phi \mbox{[Mx]}
 +\log C_A
\label{eq:A_Phi_log}
\end{equation}
with $k=0.84$ and $C_A=6.17\cdot 10^{-16}\,$, 
while $r_A$ is a Gaussian random variable with standard deviation $0.1$.

(2)  The pole separation may be best represented as 
$d=r_d \cdot \langle d\rangle$ where 
\begin{equation}
\langle d \rangle= m_d \log(\Phi/\Phi_0)
\label{eq:d_Phi_repeat}
\end{equation}
with $m_d=3{\degdot}32$ and $\Phi_0=1.6\cdot 10^{20}\,$Mx, while $\log(r_d)$ is 
a normally distributed random variable, with standard deviation 0.41.

(3) The tilt angle $\gamma_J$ is best represented as
$\langle\gamma_J\rangle + r_J \sigma_J$ where 
\begin{equation}
\langle\gamma_J\rangle= m_J \sin\lambda   \qquad m_J=28{\degdot}62
\label{eq:joylaw_repeat}
\end{equation}
while $r_J$ is a random variable with a distribution described 
by Student's $t$-distribution of $3.7$ degrees of freedom, and 
$\sigma_J$ is obtained from
\begin{eqnarray}
&&\sigma_J =\sigma_\infty + A\exp (-d/d_0) \qquad \mbox{with}
\label{eq:joyscatter_repeat} \\
  &&\sigma_\infty = 11{\degdot}0 \qquad A =
  33{\degdot}5 \qquad 
  d_0 =  4{\degdot}4 \nonumber
\end{eqnarray}
or alternately from
\begin{eqnarray}
&&\sigma_J =a+b\log(\Phi \mbox{[SFU]}) \quad \mbox{with}
\label{eq:joyscatter_Phi_repeat} \\
  &&a = 25{\degdot}57 \quad \mbox{and} \quad 
  b =  3{\degdot}75 \nonumber
\end{eqnarray} 
in which case a $t$-function with 2.9
degrees of freedom is to be used. 

For applications where an accurate representation of the tails of the
distribution matters, accounting for the skewness by suppressing
[amplifying] the positive [negative] tail of the distribution may be
considered. 

\end{appendix}

\end{document}